\documentclass[twocolumn,showpacs,preprintnumbers,amsmath,amssymb]{revtex4}
\usepackage{tabularx,graphicx}\begin{document}
\newcommand{\beq}{\begin{equation}}
\newcommand{\eeq}{\end{equation}}
\newcommand{\beqn}{\begin{eqnarray}}
\newcommand{\eeqn}{\end{eqnarray}}
\newcommand{\bmath}{\begin{subequations}}
\newcommand{\emath}{\end{subequations}}
\title{Electromotive forces and the Meissner effect puzzle}
\author{J. E. Hirsch }
\address{Department of Physics, University of California, San Diego,
La Jolla, CA 92093-0319}

\begin{abstract} 
 In a voltaic cell, positive (negative) ions flow from the low (high) potential electrode to the high (low) potential electrode, 
 driven by an `electromotive force' which points in opposite direction and overcomes the electric force.   Similarly in a superconductor charge flows in direction opposite to that
 dictated by the Faraday electric field as the magnetic field is expelled in the Meissner effect. The puzzle is the same in both cases: what drives electric charges against electromagnetic
 forces? 
 I propose that the answer is also the same in both cases:  kinetic energy lowering, or `quantum pressure'.  
   \end{abstract}
\maketitle 
\section{introduction}

   \begin{figure}
 \resizebox{6.5cm}{!}{\includegraphics[width=7cm]{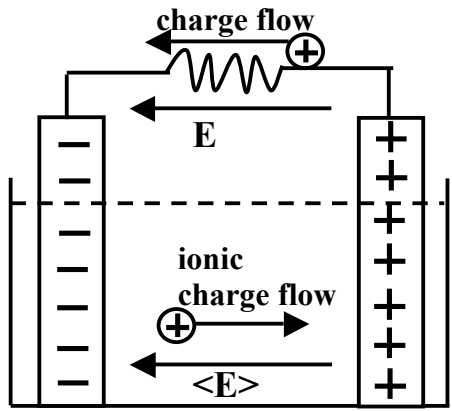}}
 \caption {In the circuit outside the voltaic cell,   positive charge flows from the positive to the negative electrode driven by an electric field E (or negative charge flows from the negative to the positive electrode).
 Inside the cell, positive (negative)  ionic charge flows in direction opposite (equal) to that of  the average electric field $<E>$, which points in the same direction as outside 
 because the electrostatic field is conservative. What drives the ionic charge flow?  }
 \label{figure2}
 \end{figure}
 
What is the force that ``pumps'' positive charges from the low potential electrode to the high potential electrode in a voltaic cell, thus creating the cell electric potential
difference that drives the circuit's electric current? (Fig. 1).  It was termed {\it electromotive force} (emf) by Volta and was the subject of much debate during the 19th century.  
Current elementary physics and chemistry textbooks will tell us that it is a `chemical force' that drives charges in direction opposite to that dictated by the electromagnetic forces, without 
discussing  it much further.
Electrochemistry  texts will give detailed explanations using oxidation and reduction potentials, contact potentials, free energies, electrochemical potentials, 
concentration gradients, etc., without clarifying  the essential physics. Both physics and chemistry texts usually will  say that the term `emf' is a misnomer, or that it is `outdated'. Instead I will argue that it is a useful and  physical concept.

 \begin{figure}
 \resizebox{8.5cm}{!}{\includegraphics[width=7cm]{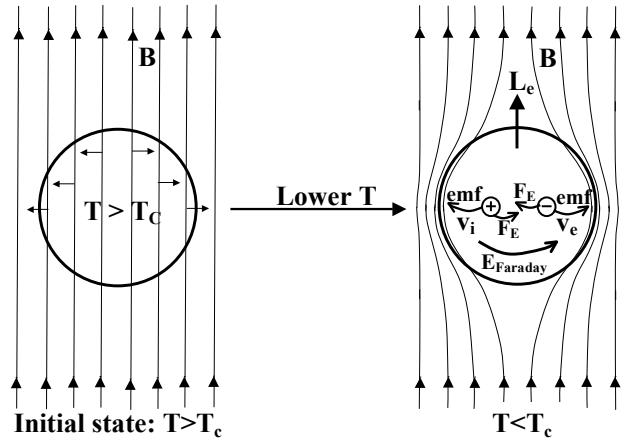}}
 \caption {When the superconductor expels a magnetic field, an electric field $E_{Faraday}$ is generated in the direction shown in the figure,
 which pushes positive and negative charges in the direction of the respective electric forces $F_E$ shown.
 However, the charges move in opposite direction,   driven by   emf's. $v_i$ and $v_e$ denote the velocity of the positive ions and negative conduction electrons, and the angular momentum of the electrons in the Meissner current is denoted by $L_e$.
   }
 \label{figure2}
 \end{figure} 
 
When a superconductor expels a magnetic field (Meissner effect) a similar puzzle arises (Figure 2). As the magnetic  flux through the superconductor decreases a Faraday electric field is
generated that exerts forces on the charges in the superconductor (negative electrons and positive ions) in the direction to create an electric current that will
restore the magnetic field in the interior (Lenz law)\cite{lenz}. However the charges in the superconductor defy these electromagnetic  forces, because the end result is  that 
the mobile negative carriers near the surface and the positive ions of the solid\cite{gyro} both end up moving in direction exactly opposite to what was prescribed by
the electromagnetic forces, so that the magnetic field in the interior of the superconductor is nullified and angular momentum is conserved\cite{missing}.

Thus, there is clearly an analogy between  the phenomena  described in the two preceding paragraphs: electric charges defying electromagnetic forces.
 There is also a difference: the electric field in the case of the voltaic cell is conservative,
and for that reason the problem is usually phrased in terms of potentials rather than forces. Instead, in the Meissner case the electric field arising from Faraday's law is non-conservative
and an electric potential cannot be defined. Nevertheless, I argue that there is an intimate connection between both situations, which is highlighted by describing
them using the concept of electromotive force rather than potentials. 
 
Strangely, the question of what is the `force' propelling the mobile charge
carriers and the ions in the superconductor to move in direction opposite to the electromagnetic force in the Meissner effect was
essentially  never raised nor answered
to my  knowledge\cite{question}.
It is generally believed that BCS theory explains the Meissner effect, however the `electromotive force' that drives charges in the superconductor
against the electromagnetic forces has not been clearly identified.
Fortunately, the question is better understood for the electromotive force in voltaic cells. Under the assumption that nature economizes on its bag of tricks,  insight  gained from the emf of voltaic cells can
  help us understand the `emf' in superconductors.

\section{the emf in a voltaic cell}

  \begin{figure}
 \resizebox{6.5cm}{!}{\includegraphics[width=7cm]{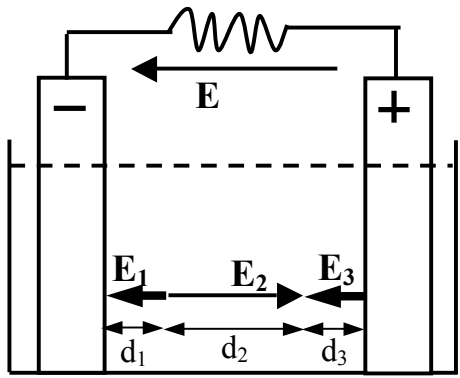}}
 \caption {  The electric field in the interior of the voltaic cell is highly inhomogeneous. It points to the left near the electrodes 
 (regions $d_1$ and $d_3$) and to the right over most of the
 extent of the cell (region $d_2$). Furthermore, $E_1$ and $E_3$ are $much$ larger than $E_2$.}
 \label{figure2}
 \end{figure} 
 
In the case of the voltaic cell, the puzzle arises from the fact that the electrostatic field is conservative:
\beq
\oint \bold{E}\cdot\bold{dl}=0 .
\eeq
Consequently, if the electric field $E$ points from the positive to the negative electrode outside the voltaic cell, it also has to point (on the average) from the positive to the negative electrode inside the cell,
as shown in Fig. 1.
However,  the charge flow carried by the ions inside the cell is in opposite direction to the charge flow by electrons in the outside circuit,
as depicted in Fig. 1, so that no charge accumulation on either electrode occurs. How do the ions manage to flow against the electric field?

  \begin{figure}
 \resizebox{6.5cm}{!}{\includegraphics[width=7cm]{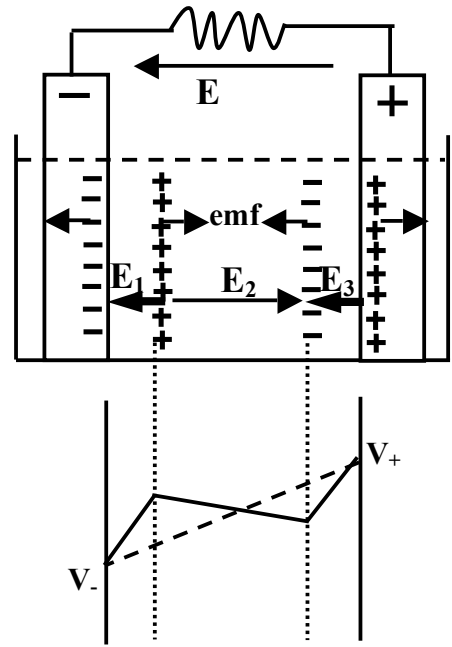}}
 \caption {A `double layer' of charges exists next to each electrode, the charges being pulled apart by the emf, giving rise to very large electric fields in those regions
 ($E_1$ and $E_3$). 
 In the bulk of the cell, the electric field $E_2$ points in direction opposite to the electric field outside ($E$) when the circuit is closed and current
 circulates. If the circuit is open   $E_2=0$. The lower part of the figure shows the electric potential inside the cell (full line) and
 outside the cell (dashed line) when the circuit is closed. The dotted lines denote the boundaries of the double layers. 
 Note that the slope of the full line between the dotted lines has opposite sign to the slope of the dashed line, corresponding to the opposite 
 directions of $E_2$ and $E$. When the circuit is open, the
 full line between the dotted lines becomes horizontal, and $E_2=0$.
   }
 \label{figure2}
 \end{figure} 

The answer is that the electric field in the interior of the cell is highly inhomogeneous,   as depicted in Figs. 3 and 4. Over most of the cell (region denoted by $d_2$ in Fig. 3)
 the electric field $E_2$ points indeed in direction opposite to
the electric field outside and drives charged ions according to the electric force. However, close to each electrode there is a layer 
(of thickness $d_1$ and $d_3$ respectively) where
an enormously larger electric field ($E_1$ and $E_3$) points in the same direction as the field outside, i.e. to the left. These electric fields satisfy
\beq
E_1 d_1 - E_2 d_2 +E_3 d_3=V_+-V_- =\Delta V >0
\eeq
where $V_+$ ($V_-$) is the electric potential of the right (left) electrode in the figures,  and $\Delta V=V_+-V_-$ is positive, so that Eq. (1) is satisfied.

The fields $E_1$ and $E_3$ exist over so-called `double layers' of several Angstrom thickness each (Figure 4), formed by ions in the solution and charges in the electrodes. Fig. 4 also shows the behavior of the electric potential, both outside (dashed line) and inside the cell (full line).

Clearly, to set up the double layers involves charges moving against electric forces, so there has to be an emf that pulls apart  positive and negative charges that attract each other through the electrostatic force, and it costs electrostatic energy. Who pays for it? Rather than `chemistry' the answer
is, of course, quantum mechanics.

\section{origin of the emf in the  voltaic cell}

 \begin{figure}
 \resizebox{6.5cm}{!}{\includegraphics[width=7cm]{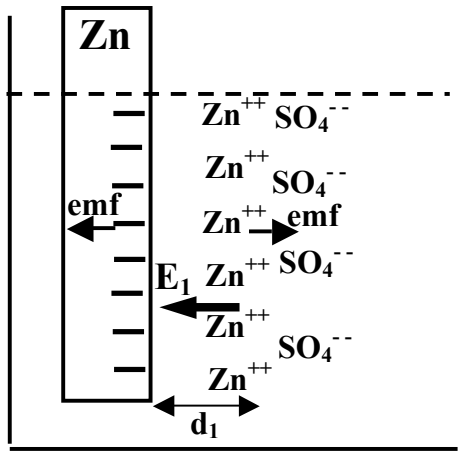}}
 \caption {Negative electrode in a voltaic cell. $Zn$ atoms of the electrode dissolve in a $SO_4Zn$ solution leaving two electrons on the electrode. The electrode is negative with respect
 to the solution, and a large electric field $E_1$ exists within a double layer of several Angstrom thickness, that exerts an electric force that tries to bring the charges back together.  
 The electromotive forces  (emf) pull in the opposite direction separating the positive and negative charges, balancing the electric force.
  }
 \label{figure2}
 \end{figure} 
 
 Figure 5 shows the negative electrode of a voltaic cell, taken to be $Zn$ for definiteness, in a $ZnSO_4$ solution. $Zn$ atoms from the electrode dissolve in the solution and
 transfer two electrons to the electrode, rendering it negative with respect to the solution. The dissolved $Zn^{++}$ ions are attracted to the electrode and remain close to it. 
 A ``double layer'' forms of several Angstrom thickness where the electric field pointing from the solution to the negative electrode is neutralized.
 
 Forming the double layer costs electrostatic energy, given by
 \beq
U_E= \int \frac{E^2}{8\pi} \sim \frac{E_1^2}{8\pi}d_1A_1 
\eeq
where $A_1$ is the cross-sectional area of the double layer of thickness $d_1$. The electrostatic force pulls in the direction of reducing $d_1$ and $U_E$, and the emf pulls in the
opposite direction.
 
 The Hamiltonian for a system of interacting charges $q_i$ in the absence of electric current is
 \beq
 H=\sum_i -\frac{\hbar^2}{2 m_i} \nabla_i^2+\sum_{i\neq j} \frac{q_i q_j}{|\bold{r}_i-\bold{r}_j|} \equiv K+U
 \eeq
 where $K$ and $U$ are kinetic and potential (Coulomb interaction) energies. The energy of the system is the expectation value of $H$, i.e. the sum of expectation values of kinetic and potential energies:
 \beq
 E=<\psi|K|\psi> + <\psi|U|\psi>
 \eeq
 It is clear that the electrostatic energy Eq. (3) arises from the expectation value of the second term in Eq. (5), the Coulomb interaction energy. Therefore, if the double layer forms
 spontaneously, the first term in Eq. (5), the expectation value of the kinetic energy, has to decrease. Since the ionic masses are much larger than the electron mass,
 the decrease in kinetic energy is dominated by the $electronic$ kinetic energy.
 
 It is easy to understand   why there is a decrease of the electronic kinetic energy as the double layer forms. The kinetic energy of a quantum-mechanical electron
 of mass $m_e$  is given by
 \beq
 \epsilon_{kin} \sim \frac{\hbar^2}{2m_e\lambda^2}
 \eeq
  where $\lambda$ measures the spatial extent where the electron is confined. When an electron is transfered from a neutral $Zn$ atom in solution to the
  metal electrode, the spatial extent of its wavefunction is no longer confined to the dimension of the single atom but rather it expands to other atoms in the
  electrode, thus lowering its kinetic energy.
  
  In conclusion I argue that formation of the double layer is driven by {\it electronic kinetic energy lowering}, or wavelength expansion, for the electrons transfered from the
  metal atom in solution to the metal electrode. This lowering of quantum kinetic energy counteracts
  and overcomes  the increase of potential (Coulomb) energy caused by charge separation, 
  hence is the origin of the emf.

The tendency of quantum-mechanical particles to $expand$ their wavefunction to lower their kinetic energy can be termed `quantum pressure', and is the most fundamental manifestation
of quantum mechanics, underlying the very stability of matter\cite{lieb}. This quantum pressure and associated quantum force is universal and acts always in a $radial$ direction.
Thus it is qualitatively different from the `quantum force' postulated by Nikulov\cite{nikulov} to explain the Little-Parks effect in superconductors. I return to this point in
a later  section.

\section{the emf in  a superconductor}
In the Meissner effect in superconductors, electric charges move in direction opposite to that dictated by the Faraday electric force, as depicted in Fig.2. Therefore, 
 a non-electromagnetic emf force is needed to explain this process, just as in the case of the voltaic cell.
 
 To visualize the emf in a superconductor undergoing the Meissner effect it is useful to think of it as a solenoid, as depicted in Fig. 6(a). A time-dependent current $I_w(t)$ flows through  a circuit with  
 a resistor $R$ and an inductor (solenoid) of self-inductance $L$ and $n$ turns per unit length, driven by a voltaic cell with emf $\epsilon$. A counter-emf $\epsilon_L$ opposes the growth of the current and the
 change in magnetic flux inside the solenoid:
 \beq
 \epsilon_L=-\frac{1}{c}\frac{\partial \phi_B}{\partial t}=-L\frac{\partial I_w}{\partial t}
 \eeq
 with 
 \beq
 \phi_B=\int \bold{B}\cdot\bold{n} dS
 \eeq
the magnetic flux through the interior of the solenoid, 
with the counter-emf $\epsilon_L$ 
 opposite to the direction of the driving emf of the battery, $\epsilon$. At any instant of time,
$\epsilon - I_wR - \epsilon_L=0$ (loop rule). The driving emf supplies the energy needed to build up the magnetic field in the interior of the solenoid
\beq
U_B=\int \frac{B^2}{8\pi} dV
\eeq
by doing work against the Faraday counter-emf $\epsilon_L$. The energy per unit time supplied by the emf is $\epsilon I_w$, 
of which $\epsilon_L I_w$ is stored in the solenoid and $I_w^2R$ is dissipated in the resistor.
As discussed in the previous section, the energy supplied by the voltaic emf originates in kinetic energy lowering,
hence the kinetic energy lowering pays the price for the magnetic energy cost Eq. (9), as well as for the thermal energy dissipated in the resistor. The self-inductance of the solenoid
$L$ is given by 
\beq
L=\frac{4 \pi^2}{c^2}n^2 h R
\eeq
with   $h$ and $R$ the height and radius of the solenoid respectively. The magnetic energy is given by
\beq
U_B=\frac{1}{2} L^2 I_w^2 =  \frac{B^2}{8\pi} V
\eeq
with $V$ the volume of the solenoid, and 
\beq
B=\frac{4\pi}{c} nI_w
\eeq
according to Ampere's law.

 \begin{figure}
 \resizebox{8.5cm}{!}{\includegraphics[width=7cm]{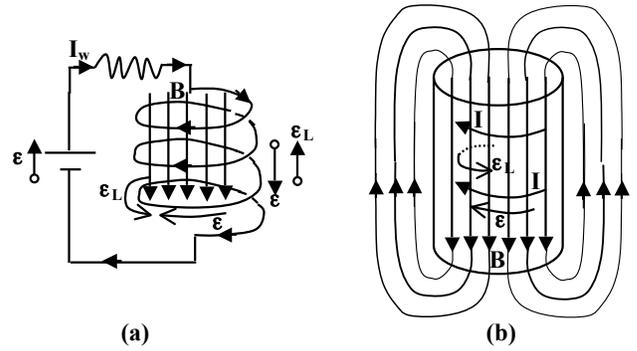}}
 \caption {In an electric circuit with a solenoid, a counter-emf 
 $\epsilon_L$ is generated
 as the current changes ((a)). The emf  $\epsilon$ does work against the 
 counter-emf, and this work is stored in the magnetic energy of the
 magnetic field that develops inside the
 solenoid. The same thing happens in a superconductor undergoing
 the Meissner effect ((b)), which develops a magnetic field $B$ of the same
 magnitude and opposite direction  as the externally applied magnetic field.
The external magnetic field is not shown in the figure.
   }
 \label{figure2}
 \end{figure}  
 
\section{origin of the emf in  the superconductor}
Similarly, we can think of the cylindrical superconductor as a solenoid,  that develops a current and
a magnetic field, as shown in Fig. 6(b), by current flowing within a London penetration depth
$\lambda_L$ of the surface. I assume there is an external magnetic field (not shown in the figure), equal and opposite to the magnetic field in the interior of the
cylinder generated by the Meissner current. The total current $I$ is related to the current in the wire of the circuit in Fig. 6(a) by $I=NI_w$, with $N$ the number of 
turns of the solenoid ($N=nh$). The current density $j$ is given by 
\beq
I=j\lambda_L h
\eeq
with $h$ the height of the cylinder. The magnetic field inside is given by
\beq
B=\frac{4 \pi}{c} \lambda_L j
\eeq
and the magnetic flux by
\beq
\phi_B=\pi R^2 B = \frac{4\pi^2}{c} R^2 \lambda_L j
\eeq
The counter-emf is given by
\beq
\epsilon_L=\oint \bold{E}\cdot \bold{dl}= -\frac{1}{c} \frac{\partial \phi_B}{\partial t}=
-\frac{4\pi^2}{c^2} R^2 \lambda_L \frac{\partial j}{\partial t}
\eeq
and the power required to drive the current $I$ against the counter-emf is
\beq
P=\epsilon_L I=\frac{4\pi}{c}\lambda_L^2 j\frac{\partial j}{\partial t} V
\eeq
with $V=\pi R^2 h$ the volume of the cylinder. Eq. (17) gives the energy per unit time
delivered to the system by the electromotive force. The total energy delivered in building
up the Meissner current is
\beq
U_B=\int_o^\infty P dt=\frac{2\pi}{c^2} \lambda_L^2 j^2 V=\frac{B^2}{8\pi} V
\eeq
as required by energy conservation.

In the voltaic cell, I have argued that the emf originates in {\it charge separation driven by
electronic kinetic energy lowering}. It is natural to expect that the emf driving the Meissner current in the
superconductor also originates in  charge separation driven by electronic  kinetic energy lowering.

Indeed, that is precisely the scenario predicted within the theory of hole superconductivity\cite{chargeexp}.
When a metal is cooled into the superconducting state, in the presence or absence of a 
magnetic field, negative charge is expelled from the interior towards the surface,  as shown schematically in Fig. 7. The driving force is kinetic energy 
lowering\cite{kinenergy}, or equivalently
wavelength expansion\cite{wavelength}: the spatial extent of the electronic wavefunction at the Fermi energy expands
from $k_F^{-1}$ to $2\lambda_L$\cite{sm}, with $k_F$ the Fermi wavevector which is of order of the inverse  interatomic distance for 
a nearly full band.
As the negative charge is expelled, it performs work against the
electric field that is created that pulls the negative charge towards the interior. In addition, in the presence of an external magnetic field
an azimuthal Meissner current is generated by the magnetic Lorentz force acting on the radially outgoing charge\cite{lenz,lorentz}, which performs work against the Faraday counter-emf that is
generated as the  magnetic field is being expelled by the Meissner current.
We  can
think of the emf as a radial force pulling  the negative charge outward against the
electric force $F_E$ that tries to maintain a uniform charge distribution and against the azimuthal Faraday counterforce 
that opposes the creation of the Meissner current when a
magnetic field is present. The expelled carriers acquire an azimuthal velocity that gives rise to a 
pure spin current in the absence of an external magnetic field through the spin-orbit interaction \cite{sm}, and to a spin current together with a charge current in the presence of an
external magnetic field through the combined action of spin-orbit interaction and Lorentz force \cite{sm,missing}.

Thus, for a metal undergoing a transition to the superconducting state in the
presence of an external magnetic field, the emf has three different tasks: (1) To deliver the kinetic energy that the carriers of the Meissner current acquire,
(2) to act against and overcome the azimuthal force resulting from the
counter-emf generated by Faraday's law, and (3) to act against   the radial electric force that opposes the negative charge
expulsion. For a metal undergoing a transition to the superconducting state in the
absence of an external magnetic field, the emf still has to provide the energy for (1) (kinetic energy of the spin current carriers) and (3) (cost in electrostatic energy of 
charge separation) (Fig. 7), however the counter-emf cost (2) is absent.

It is interesting that all  these energy costs are closely related. In the theory of hole superconductivity, 
in the absence of magnetic field the carriers near the surface acquire a spin current velocity\cite{sm}
\beq
v_\sigma ^0= \frac{\hbar}{4m_e \lambda_L}
\eeq
and have kinetic energy
\beq
K=\frac{1}{2}m_e (v_\sigma ^0)^2=\frac{\hbar^2}{32m_e \lambda_L^2}
\eeq
hence the kinetic energy density per unit volume is
\beq
u_K=n_s\frac{\hbar^2}{32m_e \lambda_L^2}=\frac{\pi}{2} (n_s \mu_B)^2
\eeq
with $\mu_B=|e|\hbar/2m_e c$ the Bohr magneton, $n_s$ the density of superconducting electrons, and the London penetration depth given by the usual expression
\beq
\frac{1}{\lambda_L^2} =  \frac{4 \pi n_s e^2}{m_e c^2} .
\eeq
The electric field inside the cylindrical superconductor (far from the surface) is
\bmath
\beq
E(r)=E_m \frac{r}{R}
\eeq
with\cite{electrospin}
\beq
E_m=-\frac{h c}{4e\lambda_L^2}
\eeq
\emath
and the average electrostatic energy density per unit volume is
\beq
u_E=\frac{E_m^2}{16\pi} =n_s\frac{\hbar^2}{64m_e \lambda_L^2}=\frac{\pi}{4} (n_s \mu_B)^2=\frac{u_K}{2}
\eeq

 \begin{figure}
 \resizebox{8.5cm}{!}{\includegraphics[width=7cm]{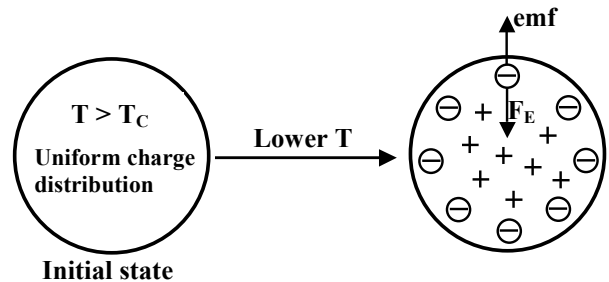}}
 \caption {According to the theory of hole superconductivity, negative charge is expelled from the interior of the superconductor towards the surface in the 
 transition to superconductivity. An outward pointing electric field is created that pulls the negative charges inward, however electrons defy this
 electric force $F_E$ as they are driven by an `emf'. 
  }
 \label{figure2}
 \end{figure}

In the presence of an external magnetic field $B$, the superfluid carriers acquire a charge velocity
\beq
v_s=-\frac{e}{m_e c}\lambda_L B
\eeq
and the increase in kinetic energy per carrier is
\beqn
\Delta K&=&\frac{1}{2}[\frac{1}{2}m_e(v_\sigma ^0+v_s)^2+\frac{1}{2}m_e(v_\sigma ^0-v_s)^2]-\frac{1}{2}m_e(v_\sigma ^0)^2  \nonumber \\
&=&\frac{1}{2}m_e v_s^2=\frac{e^2}{2m_e c^2}\lambda_L^2 B^2
\eeqn
The maximum magnetic field that the superfluid can sustain brings the motion of the carriers of spin parallel to the magnetic field to a halt\cite{sm} and doubles the 
speed of the carriers of spin antiparallel to the magnetic field. It   is
given by
\beq
B_s=-\frac{m_e c}{e\lambda_L} v_\sigma ^0=-\frac{\hbar c}{4 e \lambda_L^2}=E_m
\eeq
and for that case the increase in kinetic energy per carrier  is
\beq
\Delta K=\frac{1}{2}[\frac{1}{2}m_e (2v_\sigma^0)^2+0]- \frac{1}{2} m_e (v_\sigma ^0)^2=K
\eeq
so that the kinetic energy per carrier is doubled. The kinetic energy density for the maximum magnetic field is then
\beq
u_K({B_s})=2u_K
\eeq

In addition, in setting up the Meissner current the emf does work against the counter-emf, and the energy density cost is
\beq
u_B=\frac{B^2}{8\pi}
\eeq
For the maximum magnetic field
\beq
u_{B_s}=\frac{B_s^2}{8\pi}=2u_E=u_K
\eeq

In summary, all these energy costs are of order $u_K$, Eq. (21). The kinetic energy cost of the spin current Eq. (21) and the charge current Eq. (28) only
apply in the region within $\lambda_L$ of the surface, where these currents exist.  The  energy supplied by the emf in setting up these currents is
\beq
\Delta U_{kin}(B)=u_K(1 + (\frac{B}{B_s})^2)\frac{\lambda_L}{R}V .
\eeq
and the energy supplied by the emf to overcome electromagnetic forces, i.e.
 in setting up the charge-separated state (Fig. 7) and in supplying the counter-emf to expel the magnetic field, is
\beq
\Delta U_{em}(B)=(u_E+u_B)V=u_K(\frac{1}{2}+(\frac{B}{B_s})^2)V
\eeq

\section{`Quantum force'}
As discussed in the previous section, to create the Meissner current requires a `force' that both provides the kinetic energy acquired by the 
carriers of the Meissner current and acts against the Faraday counter-emf that is generated as the
system expels the external magnetic field. The conventional theory of superconductivity does not provide an understanding of how this azimuthal force on the
superfluid electrons near the surface is generated. Instead, it argues that because the BCS wavefunction for the superfluid in the presence of an external magnetic  field
has canonical momentum $\bold{p}=0$ and lower energy than the normal state wavefunction, the system will find its way to the superconducting state starting from
the normal state and generate the required Meissner current. However, I argue that it should be possible to understand
using classical or semiclassical concepts what are the forces on the electrons that cause them to develop the Meissner current.

The canonical momentum of the superfluid electrons is given by
\beq
\bold{p}=m_e \bold{v} +\frac{e}{c} \bold{A}
\eeq
where $\bold{A}$ is the magnetic vector potential. In the initial state before the supercurrent develops, $\bold{v}=0$ and
\beq
\bold{p}= \frac{e}{c} \bold{A}=\frac{e}{2c}\bold{B}\times\bold{r}
\eeq
in the presence of a uniform magnetic field. In the superconducting state, $\bold{p}=0$ throughout the volume of a simply connected superconductor
and the superfluid acquires  a velocity
\beq
\bold{v}=-\frac{e}{m_ec}\bold{A} .
\eeq
Hence the change in canonical momentum required is
\beq
\Delta\bold{p}=\frac{e}{2c}\bold{B}\times\bold{r}
\eeq
which points {\it in the azimuthal direction}. For a multiply connected superconductor (eg a ring) $\bold{p}$ satisfies the quantum condition
\beq
\oint\bold{p}\cdot\bold{dl}=nh
\eeq
with $n$ integer. In general, this condition will not be satisfied for an arbitrary initial $\bold{B}$ by Eq. (34), and a Meissner current will be generated
so that the magnetic flux in the interior of a superconducting ring is quantized.

Nikulov\cite{nikulov} correctly recognized that this is a fundamental unanswered question in the conventional theory of
superconductivity. He postulates the existence of an azimuthal `quantum force' $\bold{F}_q$ that acts on the superfluid electrons when the system is cooled below $T_c$, 
that forces the canonical momentum to change to satisfy the quantum condition Eq. (38)
(or equivalently  that forces the macroscopic wave function to be single-valued), given by
\beq
\bold{F}_q=\Delta \bold{p}\omega 
\eeq
with $\omega^{-1}$ the time scale over which the canonical momentum changes. This force is supposedly uniformly distributed around the loop.
Nikulov claims that this force explains the Meissner effect as well as the Little Parks effect.

However, I argue that there is no physical basis for such an azimuthal force. For one thing, we know of no other physical system where such a force
manifests itself. In addition, an azimuthal quantum force acting on electrons only would change the total angular momentum of the system, violating the physical principle of angular momentum conservation. Consequently, one would have to assume that this
`quantum force' acts on $both$ electrons and ions imparting them with equal and opposite angular momentum. 
Because we can think of the massive ions as essentially classical objects it is farfetched to assume that Nikulov's `quantum force' would 
act on them. Finally, Nikulov's quantum force exists only in the presence of a magnetic field, but no insight is provided for how 
the magnetic field would give rise to this azimuthal force.

Instead, consider the equation of motion of an electron in the presence of electric and magnetic fields
\beq
\frac{d\bold{v}}{dt}=\frac{e}{m_e}\bold{E}+\frac{e}{m_ec}\bold{v}\times\bold{B}  .
\eeq 
Using
\beq
\frac{d\bold{v}}{dt}=\frac{\partial \bold{v}}{\partial t} + (\bold{v}\cdot \bold{\nabla})\bold{v}= \frac{\partial \bold{v}}{\partial t} +\bold{\nabla}(\frac{\bold{v}^2}{2})-\bold{v}\times(\bold{\nabla}\times\bold{v})
\eeq
and Faraday's law $\bold{\nabla} \times \bold{E}=-(1/c)\partial \bold{B}/\partial t$  it follows that
\beq
\frac{\partial \bold{w}}{\partial t} = \bold{\nabla}\times(\bold{w}\times\bold{v})
\eeq
for the `generalized vorticity'
\beq
\bold{w}=\bold{\nabla}\times\bold{v}+\frac{e}{m_ec}\bold{B} .
\eeq
which is related to the canonical momentum by $\bold{w}=(\bold{\nabla}\times\bold{p})/m_e$.
In the initial  state, at time $t=0$:
\beq
\bold{w}(\bold{r},t=0)=\frac{e}{m_ec}\bold{B}(t=0)\equiv \bold{w}_0
\eeq
independent of position $\bold{r}$.  In the superconducting state, Eq. (36) is satisfied, hence
\beq
\bold{w}(\bold{r},t=\infty)=0
\eeq
In a cylindrical geometry, assuming azimuthal symmetry as well as translational symmetry along the cylinder axis ($z$) direction (infinitely long cylinder)
\beq
\bold{w}(\bold{r},t)=w(r,t)\hat{z}
\eeq
and Eq. (42) takes the form
\beq
\frac{\partial w}{\partial t} =-\frac{1}{r}\frac{\partial}{\partial r}(rwv_r)
\eeq
with $r$ the radius in cylindrical coordinates. Eq. (47)  shows that  the only way $w$ can change from its initial non-zero value to zero is if
{\it the radial velocity $v_r$ is non-zero}.
Moreover,  for $w$ to evolve towards
its final value $0$ requires $v_r>0$, i.e. a radial $outflow$ of electrons. BCS theory does not predict a radial outflow of electrons in the transition to
superconductivity, hence within Eq. (47) it predicts that $w$ does not change with time. Therefore, I argue that within
BCS theory the Meissner effect does not take place!

On the other hand, as discussed earlier the theory of hole superconductivity predicts expulsion of negative charge from the interior towards the surface as the 
system becomes superconducting. Expulsion of charge necessarily involves a radial 
velocity $v_r$, hence the theory of hole superconductivity allows for a change in $w$ through Eq. (47) as required for the Meissner effect to take place.
The magnetic Lorentz force acting on radially outgoing electrons gives rise to an azimuthal force in the direction required to generate
the Meissner current. The expulsion of negative charge can be understood as arising from a $radial$ quantum force, or `quantum pressure', as discussed in the 
next section. 

\section{quantum pressure and phase coherence}
In the theory of hole superconductivity, the transition to superconductivity can be understood as an expansion of electronic orbits from radius 
$k_F^{-1}$ to radius $2\lambda_L$\cite{sm}, as shown schematically in Fig. 8.
 \begin{figure}
 \resizebox{8.5cm}{!}{\includegraphics[width=7cm]{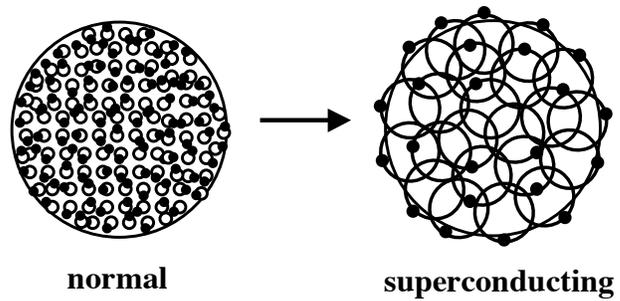}}
 \caption {Electronic orbits in the normal state have radius $k_F^{-1}$, of order of the ionic lattice spacing, and electronic orbits don't overlap.
 In the transition to superconductivity the orbits expand to radius $2\lambda_L$, several hundreds Angstrom, and they become highly overlapping.
 The black dots on the orbits indicate the `phase' of the electron, which is random in the normal state and coherent in the superconducting state.
   }
 \label{figure2}
 \end{figure} 
 This can be justified from the form of the orbital magnetic susceptibility. In the normal state it is given by the Landau diamagnetism formula
\bmath
 \beq
 \chi_{Landau}=-\frac{1}{3}\mu_B^2g(\epsilon_F)=-\frac{n_se^2}{4m_e c^2}k_F^{-2}
 \eeq
 where I have used the free electron density of states $g(\epsilon_F)=3n_s/2\epsilon_F$ and $\epsilon_F=\hbar^2k_F^2/2m_e$. In the perfectly
 diamagnetic superconducting state
  \beq
 \chi_{London}=-\frac{1}{4\pi}=-\frac{n_se^2}{4m_e c^2}(2\lambda_L)^{-2}
 \eeq
 \emath
 Both expressions Eqs. (48a) and (48b) correspond to the Larmor formula for the magnetic susceptibility of electrons of density $n_s$ in orbits perpendicular to the
 applied magnetic field with radius $k_F^{-1}$ and $(2\lambda_L)$ respectively.  
 
 Such an expansion of the orbits will lead to a decrease of the quantum kinetic energy. A lower bound for the kinetic energy of an electron described
 by the wavefunction $\psi(\bold{r})$ is\cite{lieb}
 \beq
 T_\psi \geq \frac{3}{5}    (6\pi^2)^{2/3 }\frac{\hbar^2}{2m_e} \frac{\int d^3r \rho(\bold{r})^{5/3} }{ \int d^3r \rho(\bold{r})}
 \eeq
 with $\rho(\bold{r})=|\psi(\bold{r})|^2$. For a wavefunction of spatial extent $\lambda$, $T_\psi\geq K_1  \hbar^2/(2m_e\lambda^2)$,
 with $K_1$ a constant, and as $\lambda$ increases the kinetic energy decreases rapidly.

 Why doesn't this expansion occur in the normal metallic phase? This can be understood semiclassically by considering the orbits shown in Fig. 8. 
  We can interpret the `phase' of the electronic wavefunction as the position of the electron in the circular orbit. For the highly overlapping orbits in the
 superconducting state, the phases of the different orbits need to be highly correlated to avoid collisions between electrons that would increase
 the Coulomb repulsion energy. This corresponds to the ``macroscopic phase coherence'' of the superconducting state. Instead, for the
 small non-overlapping orbits in the normal state, the phases of the different orbits do not need to be related to each other. 
 In that case there are many different ways to choose the phases of the individual orbits, consequently the normal state has much higher entropy and hence
 lower free energy at high temperatures. When the temperature is lowered enough that the lower energy of the superconducting state dominates
 the higher entropy of the normal state for the free energy $F=E-TS$, the superconducting state becomes favored and the
 orbits expand and become coherent. If a magnetic field is present, the orbit expansion leads to the Meissner effect.

\section{discussion}

In summary, I argue that: (i)  there are no $azimuthal$ `quantum forces' in nature, 
and (ii) electromagnetic forces on quantum or classical particles are described by the Lorentz force formula 
$\bold{F}=q\bold{E}+(q/c)\bold{v}\times \bold{B}$ for a particle of charge $q$\cite{lorentz}. Either (i) or (ii) (or both) have to be false  within the BCS-London conventional theory of superconductivity as well as within Nikulov's interpretation of the Meissner effect. I argue that an azimuthal `quantum force' would violate the principle of
angular momentum conservation derived from the isotropy of space, and that we know of no electromagnetic forces that do not originate in the Lorentz force,
which as I have shown cannot explain the Meissner effect in the absence of net electronic $radial$ velocity.
If (i) and (ii) are true, the Meissner effect is a fundamental unsolved puzzle within the conventional theory of superconductivity. Since the Meissner effect is the most fundamental manifestation of superconductivity,   this 
 calls the entire  validity of the conventional theory into question\cite{bcsquestion}. 

There are no azimuthal quantum forces but there $is$ a $radial$ quantum force, the radial derivative of Eq. (6):
\beq
\bold{F}_q\equiv \frac{\hbar^2}{m_e \lambda^3} \bold{\hat{r}}
\eeq
with $\lambda$ the spatial extent of the wavefunction and $\bold{\hat{r}}$ the radial direction. This force  embodies the difference between classical and
quantum physics, namely the drive of quantum particles to expand their wavefunctions to lower their kinetic energy, the more so the more confined they are
and the smaller their mass is.
This can be superficially understood using the uncertaintly principle, however as discussed e.g. by Lieb\cite{lieb}, the uncertainty principle alone is
not sufficient to explain it. Because this quantum force is radial, it does not change the angular momentum of the system as required by the principle
of angular momentum conservation, in contrast to Nikulov's  azimuthal quantum force.
 We can  understand the physical origin of this radial quantum force, or quantum pressure,
  as arising from the dependence of the classical  kinetic energy of a rotating particle of fixed
 angular momentum $\hbar$ on the radius of rotation $\lambda$: $m_e \omega\lambda^2=\hbar$ is the angular momentum for angular rotation frequency $\omega$,
 and $\epsilon_{kin}=(1/2) m_e \omega^2 \lambda^2=\hbar^2/2m_e \lambda^2$ is the kinetic energy. 

Within the theory of hole superconductivity, the Meissner effect is naturally explained as   originating in this radial ``quantum force'', 
or ``quantum pressure''. It gives rise to a {\it radial velocity}, as required for the Meissner effect by Eq. (47), and leads to the macroscopically inhomogeneous
charge distribution depicted in Fig. 7 (b) and described by the modified London electrodynamics proposed by the author\cite{chargeexp}.
The associated lowering of kinetic energy has in fact been experimentally detected in optical experiments in 
high $T_c$ cuprates\cite{basov,marel,santander}, as well as predicted by the theory of hole superconductivity\cite{apparent} well before the 
experimental detection and well before the connection between
kinetic energy lowering, charge expulsion and the Meissner effect had been elucidated. The reason the $hole$ character of the carriers in the normal state is essential
is clear: electron-like carriers in an almost empty band have already a low kinetic energy, long wavelength and a delocalized wavefunction, hence the drive to 
expand the spatial extent of the wave function to lower kinetic energy does not exist. In contrast, electrons in almost full bands have their wavefunctions compressed to
a spatial extent of order a lattice spacing and consequently highest kinetic energy.

There have been other proposals of `kinetic-energy-driven' superconductivity mechanisms\cite{kinetic,k2,k3,k4,k5,k6,k7,k8,k9,k10}.
None of them associates kinetic energy lowering with
almost full bands nor with expansion of the wavefunction nor with negative charge expulsion nor with an explanation of the Meissner effect.

In conclusion, I propose that the azimuthal force required to get the electrons in the Meissner current moving 
in the azimuthal direction and have them overcome the Faraday counter-emf
is in fact an electromagnetic force, the magnetic Lorentz force deflecting  radially outgoing electrons. The Lorentz force  transfers  azimuthal momentum to the electrons
in the Meissner current but it does not impart them with kinetic energy, since the magnetic force does not do work. The ultimate driving force for the Meissner effect, the ``emf'', 
is not electromagnetic but is the radial quantum force arising from quantum kinetic energy lowering, just as in the case of the  voltaic cell.

\end{document}